\documentstyle[aps,prd,epsf]{revtex}

\begin{document}

\title{Uncertainty for spin systems}
\author{Nuno Barros e S\'a\thanks{Email address: 
nunosa@vanosf.physto.se. Supported by grant PRODEP-Ac\c c\~ao 5.2.}}
\address{Fysikum, Stockholms Universitet, Box 6730, 
113 85 Stockholm, Sverige\\
{\rm and} DCTD, Universidade dos A\c cores, 
9500 Ponta Delgada, Portugal}
\date{\today}
\maketitle

\begin{abstract}
A modified definition of quantum mechanical uncertainty $\Delta$ for spin 
systems, which is invariant under the action of $SU(2)$, is 
suggested. Its range is shown to be $\hbar^2j\leq\Delta\leq\hbar^2j(j+1)$ 
within any irreducible representation $j$ of $SU(2)$ 
and its mean value in Hilbert space computed using the Fubini-Study 
metric is determined to be ${\rm mean}(\Delta )=\hbar^2j(j+1/2)$. 
The most used sets of coherent states in spin systems coincide with 
the set of minimum $\Delta$ uncertainty states. \end{abstract}

Coherent states are an important tool in the study of wave 
phenomena finding many relevant applications in Quantum 
physics \cite{gl1,ksu}. 
The familiar Glauber states \cite{schr,gl2} can be 
equivalently defined as the elements of the orbit of the 
Heisenberg-Weyl group which contains the ground state, as the eigenstates 
of the annihilation operator or as the minimum uncertainty 
wave-packets. Following these different definitions there are 
different approaches to the generalization of the concept of 
coherent states, the one based on the first 
definition \cite{kl1} being the most popular. 
The generalization procedure 
has been extended to include spin systems \cite{kl2,rad} and 
others \cite{kl3,bgi,per,bbd}. A full account of applications 
of coherent states in different areas of Physics can be found in 
\cite{ksk}. 

In the group theoretical approach to coherent states 
Hilbert space is decomposed into the union of disjoint sets of 
coherent states, the group orbits. For spin systems the 
orbit space (the set of orbits) is composed almost entirely of 
$3$-dimensional orbits with the exception of a finite number of 
$2$-dimensional orbits which consist of 
the eigenvectors of $\vec r\cdot\vec J$, with $\vec J$ the generators 
of the Lie algebra of $SU(2)$ and $\vec r$ any numeric vector 
\cite{bac,sa}. In each irreducible representation of $SU(2)$ there is one 
particular orbit which admits an analytic representation in the complex 
plane, 
$$|z>=\frac{1}{(1+|z|^2)^j}e^{zJ_-}|j>\ .$$
It turns out that this orbit is 
singled out by the structure of orbit space: it is the 
$2$-dimensional orbit composed of the eigenvectors of $\vec r\cdot\vec J$ 
with the highest absolute value of its eigenvalue \cite{sa}. 
However the states belonging to this orbit are not all minimum 
uncertainty states; they do not even have constant uncertainty. 

Uncertainty is an important property of a physical state, and it 
would be desirable to keep it playing a major role in the 
definition of coherent states. 
Minimum uncertainty states have been studied in \cite{nie} 
and, in the context of spin systems, states saturating the equality 
in the Heisenberg relation have been studied in \cite{acs} and 
called intelligent states. States saturating the equality in the 
Robertson relation have also been studied \cite{tri,bba}. 
One unsatisfactory feature of intelligent states and of the commonly used 
definition of uncertainty for spin systems is that they are not invariants 
under the action of $SU(2)$. As a consequence sets of coherent states based 
on these definitions cannot be represented as orbits of $SU(2)$. This is 
in contrast with the situation in particle mechanics where the 
Heisenberg inequality and the uncertainty function used 
are invariants under the action of the Heisenberg-Weyl group. 

Here we propose a new definition of uncertainty for spin systems, 
$$\Delta=\Delta {J_x}^2+\Delta {J_y}^2+\Delta {J_z}^2\ ,$$
which is a positive 
increasing function of the variances and which is invariant under the 
action of $SU(2)$. It obeys the following invariant inequalities: 
$$\hbar^2j\leq\Delta\leq\hbar^2j(j+1)\ ,$$
which play the role of uncertainty relations. As an immediate application 
we show that the particular set of coherent states which admits an analytic 
representation in the complex plane coincides with the set of minimum 
uncertainty states for this inequality. We use the Fubini-Study metric to 
compute the mean value of the uncertainty $\Delta$ in Hilbert space with the result: 
$${\rm mean}(\Delta )=\hbar^2j\left( j+\frac{1}{2}\right)\ ,$$ 
for any irreducible representation $j$. This shows that in higher 
dimensional representation spaces of $SU(2)$ most of the states have 
high  values of uncertainty. In particular one has 
\begin{equation}
\lim_{j\to\infty}\frac{{\rm mean}(\Delta )}{\max(\Delta )}=1\ .
\end{equation}

The paper is organized as follows: 
In section \ref{se1} we review some mathematical definitions concerning 
group orbits and invariants, the Glauber coherent states and their 
generalization, and the construction of spin coherent states. 
In section \ref{se2} we discuss the issue of Heisenberg-like 
inequalities and uncertainty relations. We propose the new definition 
of uncertainty $\Delta$ for spin systems and we state and prove the 
statements about $\Delta$ made above. 
We include an appendix on how to 
average quantities in $CP^N$ using the Fubini-Study metric.

\section{Introduction}\label{se1}

\subsection{Group orbits and invariants}\label{se11}

Let $U(g)$ be a representation of the Lie 
group $G$ on the Hilbert space $\cal H$. The 
$G$-orbit through $|\phi >\in{\cal H}$ is the subset of $\cal H$ given by 
\begin{equation}
{\cal C}_\phi =\left\{ |\psi >\in {\cal H}: |\psi >=
U(g)|\phi >\ ,\ g\in G\right\}\ .\label{gor}
\end{equation}
It follows that 
\begin{equation}
\dim{\cal C}_\phi\leq\dim G\ \ {\rm and}\ \ 
\dim{\cal C}_\phi\leq\dim{\cal H}\ .
\end{equation}
The relation ``$|\phi '>$ lies on the same orbit as $|\phi >$'' 
is clearly an equivalence relation: 
reflexive, symmetric and transitive. As a consequence $\cal H$ can 
be partitioned into disjoint orbits 
\begin{equation}
{\cal H}=\bigcup_\phi {\cal C}_\phi 
\end{equation}
where the label $\phi$ runs over orbits (equivalence classes) and 
not over vectors. 
The quotient space ${\cal H}/G$ is called the 
orbit space. 
A function $f(|\psi >)$ in Hilbert space $\cal H$ is said 
to be $G$-invariant if 
\begin{equation}
f(U(g)|\psi >)=f(|\psi >)\ ,\ \forall g\in G\ ,\ \forall |\psi >\in 
{\cal H}\ .
\end{equation}
It follows that $G$-invariant functions are also functions on orbit 
space ${\cal H}/G$. 

For more information on these issues see for instance \cite{mic,asa}.

\subsection{Glauber states}\label{se12}

The familiar Glauber states $|q,p>$ in particle mechanics can be seen as the 
$G$-orbit of the Heisenberg-Weyl group through the vacuum state $|0>$, 
\begin{equation}
|q,p>=U(q,p)|0>\ ,\label{rqr3}
\end{equation}
where $U(p,q)$ is the Weyl operator  
\begin{equation}
U(q,p)=e^{i(pQ-qP)/\hbar}\ .
\end{equation}
They are eigenstates of the annihilation operator and they admit 
the useful analytic representation in the complex plane 
\begin{equation}
|p,q>=e^{(za^+-z^*a)}|0>=e^{-|z|^2/2}\sum_n\frac{z^n}{\sqrt{n!}}|n>\ ,
\label{coe}
\end{equation}
with $z=(q+ip)/\sqrt{2\hbar}$. 
It can be shown that the Glauber states are minimum 
uncertainty states since 
\begin{equation}
\Delta Q^2=\Delta P^2=\hbar /2\ ,
\end{equation}
and the equality sign is satisfied in the Heisenberg uncertainty 
relation (sometimes the square root of this relation is used; here 
we prefer this form)
\begin{equation}
\Delta Q^2\Delta P^2\geq\hbar^2/4\ .\label{hei} 
\end{equation}
The remaining $G$-orbits of the Heisenberg-Weyl group can be seen as 
generalized coherent states \cite{kl1,ksk} but they are not eigenstates of any 
particularly simple operator, they do not admit an analytic representation 
in the complex plane, and they are not minimum uncertainty states. 
Nevertheless they have constant values of uncertainty since both factors 
$\Delta Q^2$ and $\Delta P^2$ are $G$-invariant functions \cite{sa}.

\subsection{Spin coherent states}\label{se13}

The group $SU(2)$ admits representations 
classified according to integer and semi-integer values $j$ with 
the Casimir operator $J^2=j(j+1)\hbar^2$. 
Let $\cal H$ be a Hilbert space carrying one such 
representation. 
Sets of generalized coherent states can be generated as the 
orbits of $SU(2)$ in $\cal H$,  
\begin{eqnarray}
{\cal C}_\phi &=&\left\{ |\vec r>\in {\cal H}: 
|\vec r>=U(\vec r)|\phi >\ ,\ \vec r\in (4\pi )^3\right\}\label{iuu}\\
U(\vec r)&=&e^{i\vec r\cdot\vec J/\hbar}\ ,\label{rqr2}
\end{eqnarray}
where we used the so-called canonical group coordinates for 
generality. 

Using the group parameterization 
\begin{equation}
U(z,\theta)=Ne^{zJ_-/\hbar}e^{-z^*J_+/\hbar}e^{-i\theta J_z/\hbar}\ ,
\end{equation}
where $J_\pm$ are the ladder operators $J_\pm =J_x\pm iJ_y$, 
and choosing the fiducial state $|\phi>$ to be an eigenstate of $J_z$, 
$|m>$ with $m=-j,..,j$, one has \cite{rad} 
\begin{equation}
|z;m>=U(z)|m>=Ne^{zJ_-/\hbar}e^{-z^*J_+/\hbar}|m>\ ,\label{an}
\end{equation}
where the phase factor resulting from $e^{-i\theta J_z/\hbar}$ 
has been ignored and $N$ stands for a normalization factor. 
Further choosing $|j>$ as the fiducial state one has
$e^{-z^*J_+/\hbar}|j>=|j>$ and 
\begin{equation}
|z>=\frac{1}{(1+|z|^2)^j}e^{zJ_-}|j>\ ,\label{anal}
\end{equation}
after determination of the normalization factor. 
This analytic representation is not available in general for the 
sets (\ref{rqr2}) generated from arbitrary fiducial vectors. 

The analogous relation for spin systems to the Heisenberg 
inequality for canonically conjugate operators (\ref{hei}) is 
\begin{equation}
\Delta J_x{^2}\Delta J_y{^2}\geq \frac{\hbar^2}{4}\overline{J_z}^2\ .
\label{var4}
\end{equation}
Notice the important difference with (\ref{hei}) that now 
the right hand side of the inequality 
is not a constant. Following \cite{acs} we shall call the 
left hand side of (\ref{var4}) the uncertainty 
$\Delta J_x{^2}\Delta J_y{^2}$. Then it is clear that the set of 
states for which the equality in (\ref{var4}) is saturated and the 
set of states of minimum uncertainty are not the same. Moreover none 
of them coincide with any set of coherent states (\ref{iuu}). On the other hand 
in particle mechanics the Glauber states satisfy the Heisenberg 
inequality and they are states of minimum uncertainty. 
In \cite{acs} the spin states satisfying the equality sign in (\ref{var4}) 
have been called intelligent states. They are given by 
\begin{eqnarray}
|\tau,N>&=&\frac{A_N}{(1+|\tau|^2)^j}\sum_{l=0}^N\left(
\begin{array}{c} N\\ l\end{array}\right) (2j-l)!\times\nonumber\\
&&\times\left(-
\frac{2}{\hbar}\tau J_+\right) ^le^{\tau J_+/\hbar}|-j>
\end{eqnarray}
where $N$ is a discrete label satisfying $0\leq N\leq 2j$ and 
$\tau$ is a continuous label which can be either real or purely 
imaginary. $A_N$ is a normalization factor. 

Finally we comment that the space of physical states for the irreducible 
representation $j$ of $SU(2)$ is $CP^N$ with $N=2j$ (see the appendix):
\begin{equation}
j\ \rightarrow\ \dim{\cal H}=2j+1\ \rightarrow\ {\rm projective\ 
space:}\ CP^{2j}\ .
\end{equation}
Its real dimension is $4j$.

\section{Uncertainty}\label{se2}

\subsection{Uncertainty relations}\label{se21}

We recall the inequality valid for hermitian operators $A$ and $B$ \cite{sha} 
\begin{equation}
\Delta A^2\Delta B^2\geq\frac{1}{4}\left( \sigma_{AB}{^2}-
\overline{[A,B]}^2\right)\ ,\label{rob}
\label{var2}
\end{equation}
where $\Delta A$ and $\Delta B$ are the standard deviations 
of the operators $A$ and $B$ 
\begin{equation}
\Delta A^2=\overline{A^2}-\overline{A}^2=
<\psi |A^2|\psi >-<\psi |A|\psi >^2\ .
\end{equation}
and where 
\begin{equation}
\sigma_{AB}=\overline{\{ A,B\} }-2\overline{A}\ \overline{B}\geq 0 
\end{equation}
is the covariance of $A$ and $B$. Since for hermitian operators 
$\sigma_{AB}$ is real and $\overline{[A,B]}$ is purely imaginary, 
both parcels on the right hand side of (\ref{var2}) are positive 
and one can state that
\begin{equation}
\Delta A^2\Delta B^2\geq -\frac{1}{4}\overline{[A,B]}^2\ .
\label{var3}
\end{equation}
This is called the Heisenberg relation while (\ref{rob}) is often called 
the Robertson relation. For canonically conjugate operators $Q$ and $P$ one 
has $[Q,P]=i\hbar$ and the Heisenberg uncertainty relation 
(\ref{hei}) follows immediately from (\ref{var3}). For spin systems 
(\ref{var4}) follows from $[J_x,J_y]=i\hbar J_z$. Notice that the 
equality can hold only if $\sigma_{AB}=0$. 

The left hand side of the Heisenberg inequality 
(\ref{var3}) is sometimes called the uncertainty. It is 
invariant under the action of the Heisenberg-Weyl group. And the right hand 
side of (\ref{var3}) is a constant. It is therefore natural to 
assign a particular physical significance to $\Delta Q^2\Delta P^2$ and 
to the states satisfying the equality sign in this inequality. 
But the left hand side of the analogous spin inequality (\ref{var4}) 
is not invariant under the action of $SU(2)$ neither is its right 
hand side a constant. Therefore there seems to be no reason why 
$\Delta {J_x}^2\Delta {J_y}^2$ should play a role for spin 
systems similar 
to the one played by $\Delta Q^2\Delta P^2$ in particle mechanics, nor why 
states saturating the equality in (\ref{var3}) or in (\ref{rob}) should be 
particularly distinguished. Such states (intelligent states) have 
been studied in \cite{acs} and in \cite{tri,bba} respectively and 
may certainly be important for the 
study of spin systems with Hamiltonians that break the $SU(2)$ 
symmetry such as systems under the action of one particular 
magnetic field pointing in the $z$-direction, but in what concerns 
the study of $CP^N$ as the representation space for spin systems 
prior to the definition of the Hamiltonian one should look for a $G$-invariant 
definition of uncertainty. We look for an uncertainty function 
which is positive and which increases with increasing values of the 
variances of the elements of the Lie algebra. The following additive 
rather than multiplicative combination of variances does the job 
\begin{equation}
\Delta =\Delta {J_x}^2+\Delta {J_y}^2+\Delta {J_z}^2\ .
\end{equation}

The following results hold:
\medskip

\noindent\underline{I} -
{\it The uncertainty $\Delta$ is $G$-invariant and therefore it is constant 
within sets of coherent states generated as orbits of $SU(2)$ in $CP^N$.} 
\medskip

\noindent\underline{II} -
{\it The uncertainty $\Delta$ is bounded from below and from above 
\begin{equation}
\hbar^2j\le\Delta\le\hbar^2j(j+1)\ .
\end{equation}
All values within this range are present in Hilbert space except 
for the representation $j=1/2$ where all states have the same 
uncertainty $\Delta=\hbar^2j$.}
\medskip

\noindent\underline{III} -
{\it The set 
\begin{equation}
\{|\psi>\in{\cal H}:\Delta(|\psi>)=\hbar^2j\}
\end{equation}
of minimum uncertainty 
vectors in the irreducible representation $j$ of $SU(2)$ coincides 
with the set of coherent states
\begin{equation}
|z>=(1+|z|^2)^{-j}e^{zJ_-}|j>
\end{equation}
generated as an orbit of $SU(2)$ in $\cal H$ and admitting an 
analytic representation in the complex plane.}
\medskip

\noindent\underline{IV} -
{\it The mean value evaluated with the volume element naturally 
associated to the Fubini-Study metric of uncertainty on the whole of 
Hilbert space is given by
\begin{equation}
{\rm mean}(\Delta)=\hbar^2j(j+1/2)
\end{equation}
for any irreducible representation $j$ of $SU(2)$.}
\medskip

Notice that the last statement is consistent with the second one for 
the $j=1/2$ representation.

\subsection{Proof}\label{se22}

\underline{I}. We have 
\begin{equation}
U^+(\vec r)J_iU(\vec r)=\Lambda_i{^j}(\vec r)J_j\ ,
\end{equation}
where $\Lambda_i{^j}$ are the matrices of the adjoint representation 
of $SU(2)$, satisfying 
\begin{equation}
\Lambda_i{^j}(\vec r)\Lambda_i{^k}(\vec r)=\delta^{jk}\ \ ,\ \forall\vec r\ .
\end{equation}
The mean values of $J_i$ transforms, within an orbit, according 
to the adjoint representation too, 
\begin{equation}
\overline{J_i}=<\vec r|J_i|\vec r>=<\phi |U^+(\vec r)J_iU(\vec r)|
\phi >=\Lambda_i{^j}(\vec r)<\phi |J_j|\phi >\ .
\end{equation}
Then $\overline{J_i}\ \overline{J_i}$ is a $G$-invariant function 
\begin{equation}
\overline{J_i}\ \overline{J_i}=\Lambda_i{^j}(\vec r)<\phi |J_j|\phi >
\Lambda_i{^k}(\vec r)<\phi |J_k|\phi >=
<\phi |J_i|\phi ><\phi |J_i|\phi >\ .
\end{equation}
This is one example of a wider set of invariants defined in \cite{sa}. 
The Casimir operator $J_iJ_i$ is invariant within the whole 
representation and consequently $\overline{J_iJ_i}$ is $G$-invariant. 
Then 
\begin{equation}
\Delta =\sum_{i=x,y,z}{\Delta J_i}^2=
\sum_{i=x,y,z}\overline{J_iJ_i}-
\overline{J_i}\ \overline{J_i}\label{uncr}
\end{equation}
is the difference between two $G$-invariant functions and is therefore 
$G$-invariant too. 
\smallskip

\underline{II}. It is always possible to choose a representative 
$|\psi>=\sum_{m=-j}^mc_m|m>$ within each orbit such that 
$<\psi|\vec J|\psi>=\overline{J_z}\vec e_z$. 
Then $\overline{J_i}\ \overline{J_i}=\overline{J_z}^2$. But 
\begin{equation}
\overline{J_z}=\sum_{m=-j}^jm\hbar |c_m|^2\ \Rightarrow\ 
|\overline{J_z}|\le\hbar j\ .\label{joz}
\end{equation}
Therefore $\overline{J_i}\ \overline{J_i}\le\hbar^2j^2$, and this 
inequality is valid all over Hilbert space since it concerns a 
$G$-invariant function. On the other 
hand it is obvious that $\overline{J_i}\ \overline{J_i}\ge 0$. 
Since $J_iJ_i=\hbar^2j(j+1)$ it follows that 
\begin{equation}
0\le\overline{J_i}\ \overline{J_i}\le\hbar^2j^2\ \Leftrightarrow\ 
\hbar^2j\le\Delta\le\hbar^2j(j+1)\ .
\end{equation}
Now we consider the one-parameter set of vectors
\begin{equation}
|\alpha>=\cos\alpha|j>+\sin\alpha|-j>\ {\rm with}\ \alpha\in 
[0,\pi/2]\ .
\end{equation}
We have
\begin{equation}
\overline{J_x}=\hbar\sqrt{j/2}\sin(2\alpha)\delta_j^{1-j}\ ,\ 
\overline{J_y}=0\ ,\ \overline{J_z}=\hbar j\cos(2\alpha)\ 
\Rightarrow\ \overline{J_i}\ \overline{J_i}=\left\{ \begin{array}{l}
\hbar^2j^2\ {\rm for}\ j=1/2\\
\hbar^2j^2\cos^2(2\alpha)\ {\rm for}\ j\ne 1/2\end{array}\right.\ .
\end{equation}
There is only one orbit in the $j=1/2$ representation \cite{sa}; 
since $\Delta$ is $G$-invariant it can only assume the value 
$\hbar^2j^2$. On the other hand, for $j\ne 1/2$ 
it is clear that $\overline{J_i}\ \overline{J_i}$ 
maps $\alpha$ onto $[0,\hbar^2j^2]$, and the statements about the 
range of $\overline{J_i}\ \overline{J_i}$ in Hilbert space are 
proven. 
\smallskip

\underline{III}. We notice from (\ref{joz}) that the 
maximum value of $\overline{J_i}\ \overline{J_i}$ is attained 
only at the vectors $|j>$ and $|-j>$ which we know to belong to 
the same orbit \cite{sa}. This single orbit coincides with the set 
(\ref{anal}) of coherent states $|z>$ since for $z=0$ we have 
$|z>=|j>$. 
\smallskip

\underline{IV}. We use the coordinates (\ref{dsoo}) defined 
in the appendix to label physical states 
\begin{equation} 
|\psi>=\sum_{m=-j}^jc_m|m>=\sum_{n=0}^NZ_n(\theta_i,\beta_j)|n-N/2>
=|\{\theta_i\} ,\{\beta_j\} >\ .
\end{equation}
Using the standard representation of the generator $J_z$ of the 
$SU(2)$ Lie algebra \cite{sak} its mean value on a state 
$|\{\theta_i\} ,\{\beta_j\} >$ is 
\begin{equation}
\overline{J_z}=\sum_{m=-j}^j|c_m|^2\hbar m=\hbar \sum_{n=0}^{N}x_n^2
\left( n-\frac{N}{2}\right) \\
\end{equation}
The mean value of $\overline{J_z}^2$ in the whole of Hilbert space 
is thus (see (\ref{av}) in the appendix) 
\begin{equation}
{\rm mean}(\overline{J_z}^2)=\frac{\hbar^2}{V_N}\int_{CP^N}dv
\overline{J_z}^2=\frac{\hbar^2}{V_N}\sum_{m,n=0}^N\left[ \left( m-
\frac{N}{2}\right)\left( n-\frac{N}{2}\right)\int_{CP^N}dv 
(x_mx_n)^2\right]
\end{equation}
Now we compute 
\begin{equation}
\int_{CP^N}dv(x_mx_n)^2=\frac{\pi^N}{(N+2)!}(1+\delta_{mn})
\end{equation}
and
\begin{equation}
\sum_{m,n=0}^N\left( m-\frac{N}{2}\right) 
\left( n-\frac{N}{2}\right)(1+\delta_{mn})=
\sum_{n=0}^N\left( n-\frac{N}{2}\right) ^2=
\frac{N(N+1)(N+2)}{12}
\end{equation}
to arrive at 
\begin{equation}
{\rm mean}(\overline{J_z}^2)=\frac{\hbar^2}{V_N}\frac{\pi^N}{(N+2)!}
\frac{N(N+1)(N+2)}{12}=\frac{\hbar^2N}{12}
\end{equation}
By symmetry one has 
\begin{equation}
{\rm mean}(\overline{J_x}^2)={\rm mean}(\overline{J_y}^2)={\rm mean}(\overline{J_z}^2)
\end{equation}
and consequently
\begin{equation}
{\rm mean}(\overline{J_i}\ \overline{J_i})=3\ 
{\rm mean}(\overline{J_z}^2)=\frac{\hbar^2N}{4}=\frac{\hbar^2j}{2}\ .
\end{equation}
The mean value of uncertainty (\ref{uncr}) in Hilbert space is 
therefore 
\begin{equation}
{\rm mean}(\Delta)={\rm mean}(\overline{J_iJ_i})-
{\rm mean}(\overline{J_i}\ \overline{J_i})=\hbar^2j(j+1)-
\frac{\hbar^2j}{2}=\hbar^2j\left( j+\frac{1}{2}\right)\ .
\end{equation}

\appendix

\section{The Fubini-Study metric and the volume element in $CP^N$}
\label{aaa}

Two vectors in Hilbert space $\cal H$ differing by a multiplicative 
non-zero complex constant $\alpha$ represent the same physical 
state, 
\begin{equation}
|z'>\sim |z>\ \ {\rm if}\ \ |z'>=\alpha |z>\label{proj}
\end{equation}
Therefore the space of physical states is the 
space of rays in Hilbert space or projective space, 
that is the space of equivalence 
classes defined by (\ref{proj}) and excluding the vector $|\psi>=0$. 
The projective spaces constructed from finite-dimensional Hilbert 
spaces are called $CP^N$ and are well studied spaces \cite{ben,kno}. The 
superscript $N$ stands for their complex dimension which is one unit 
lower then the complex dimension of the Hilbert space from which 
they are constructed. 

If $|n>$ is a basis for $(N+1)$-dimensional Hilbert space any vector 
$|\psi>$ can be written as
\begin{equation}
|\psi>=\sum_{n=0}^NZ_n|n>\ .
\end{equation}
The complex numbers $Z_n$ are homogeneous coordinates in $\cal H$ 
and they can also be used as coordinates in $CP^N$ provided one 
makes the identifications 
\begin{equation}
Z'_n\sim Z_n\ {\rm if}\ \exists\alpha:\forall n, Z'_n=\alpha Z_n\ .
\end{equation}

We start by reminding the reader that the unit $N$-sphere can be 
defined as the hyper-surface in 
$(N+1)$-dimensional Euclidean space with coordinates 
$x_i\ ,\ i=0,..,N$ that satisfies 
\begin{equation}
\sum_{i=0}^Nx_i^2=1\ .
\end{equation}
Intrinsic coordinates $\theta_i$ can be defined by 
\begin{equation}
x_i=\cos\theta_i\prod_{j=i+1}^N\sin\theta_j\ ,\label{esfcor}
\end{equation}
Their range is $(0,\pi)$ except 
for for $\theta_1$ with range $(0,2\pi)$. $\theta_0=0$ is not a 
coordinate.
The metric induced on the $N$-sphere by its embedding in 
$(N+1)$-dimensional Euclidean space in this coordinates is diagonal 
with components
\begin{equation}
g_{ii}=\left( \prod_{j=i+1}^{N}\sin\theta_j\right) ^2\ ,\label{lee}
\end{equation}
and the volume element is
\begin{equation}
dv=\prod_{i=1}^{N}\sin^{(n-1)}\theta_i\ d\theta_i\ .\label{dvx}
\end{equation}
Real projective space $RP^N$ follows the same construction with the 
range of $\theta_1$ being $(0,\pi)$ too, plus the identifications
\begin{equation}
(0,\theta_2,..,\theta_N)\equiv (\pi,\pi-\theta_2,..,\pi-\theta_N)\ .
\end{equation}

For quantum mechanical purposes the metric of interest in $CP^N$ 
is the Fubini-Study metric \cite{kno}. Its line element in 
the homogeneous coordinates $Z_i$ is 
\begin{equation}
ds^2=\frac{1}{X^2}\sum_{i=0}^{N}dZ_id\bar Z_i-\frac{1}{X^4}
\sum_{i=0}^{N}dZ_i\bar Z_i\sum_{j=0}^{N}Z_jd\bar Z_j\ ,
\end{equation}
where we have defined
\begin{equation}
X^2=\sum_{i=0}^{N}Z_i\bar Z_i\ .
\end{equation}
Splitting the complex homogeneous coordinates into their absolute 
values and phases
\begin{equation}
Z_i=X_ie^{i\alpha_i}\ ,
\end{equation}
the Fubini-Study metric splits into two blocks relative to the 
$X_i$ and to the $\alpha_i$,
\begin{equation}
ds^2=ds_X^2+ds_\alpha^2\ ,
\end{equation}
with
\begin{eqnarray}
ds_X^2&=&\frac{1}{X^2}\left( \sum_{i=0}^NdX_i^2-dX^2\right)
\label{le1}\\
ds_\alpha^2&=&\frac{1}{X^2}\sum_{i=0}^NX_i^2d\alpha_i^2-
\frac{1}{X^4}
\left( \sum_{i=0}^Nx_i^2d\alpha_i\right) ^2\label{le2}
\end{eqnarray}
The intrinsic coordinates on the sphere (\ref{esfcor}) and the 
phases relative to $\alpha_0$ 
\begin{equation}
\beta_i=\alpha_i-\alpha_0\ ,\quad i=1,..,N\label{ancor}
\end{equation}
can be used as intrinsic coordinates on $CP^N$. However we should 
remark that the ranges of all the coordinates $\theta_i$ are 
$(0,\pi /2)$ since the $X_i$ are absolute values and cannot 
therefore be negative. Moreover these coordinates are clearly 
singular whenever $\theta_i=\{ 0,\pi/2\}$. The 
relation of this coordinates with the homogeneous ones is 
\begin{equation}
Z_i=Xe^{i\alpha_0}x_i(\theta_j)e^{i\beta_i}\ .\label{dsoo}
\end{equation}
Plugging this expression into the previous formulas for the line 
elements (\ref{le1})-(\ref{le2}) one gets 
\begin{eqnarray}
ds_X^2&=&\sum_{i=0}^Ndx_i^2=\sum_{i=1}^Ng_{ii}d\theta_i^2\\
ds_\alpha^2&=&\sum_{i=1}^Nx_i^2d\beta_i^2-
\left( \sum_{i=1}^Nx_i^2d\beta_i\right) ^2
=\sum_{i,j=1}^Nh_{ij}d\beta_id\beta_j\ .
\end{eqnarray}
The first is the line element in the unit sphere (\ref{lee}) and 
in the phase line element $ds_\alpha^2$ we have defined the metric 
\begin{equation}
h_{ij}=x_i^2(\delta_{ij}-x_j^2)
\end{equation}
with inverse
\begin{equation}
h^{ij}=\frac{1}{x_0^2}+\frac{\delta_{ij}}{x_i^2}\ .
\end{equation}

The volume element for the phase coordinates is 
\begin{eqnarray}
dv_\alpha&=&\sqrt{\det (h_{ij})}\prod_{k=1}^Nd\beta_k=
\sqrt{\det (\delta_{ij}-x_j^2)}\prod_{k=1}^Nx_kd\beta_k=
\sqrt{1-\sum_{i=1}^Nx_i^2}\prod_{k=1}^Nx_kd\beta_k=
\nonumber\\
&=&\prod_{i=0}^Nx_i\prod_{j=1}^Nd\beta_j
=\prod_{i=1}^N\cos\theta_i\sin^i\theta_i\ d\beta_i\ ,
\end{eqnarray}
where we used (\ref{esfcor}) for $x_i$ in the last equality. 
Using (\ref{dvx}) for $dv_X$ the combined volume element is 
\begin{equation}
dv=dv_Xdv_\alpha=\prod_{i=1}^N\cos\theta_i\sin^{2i-1}\theta_i\ 
d\theta_id\beta_i\ .\label{vefs}
\end{equation}
The total volume of $CP^N$ becomes easy to compute
\begin{equation}
V_N=\prod_{i=1}^N\int_0^{\pi/2}d\theta_i\cos\theta_i\sin^{2i-1}
\theta_i\int_0^{2\pi}d\beta_i
=\prod_{i=1}^N\frac{1}{2i}2\pi=\frac{\pi^N}{N!}\ .\label{vcp}
\end{equation}

Now we are able to compute mean values of functions in Hilbert space as 
their integral in $CP^N$ weighted with the Fubini-Study volume 
element (\ref{vefs}) and divided by the volume $V_N$ of $CP^N$ (\ref{vcp}). 
Since the functions we are interested in are 
of the type $<\psi|A|\psi>=\overline{A}$ we shall write explicitly 
${\rm mean}(\overline{A})$ to emphasize that the mean value is not taken 
on quantum states but rather on the whole of $CP^N$, 
\begin{equation}
{\rm mean}(\overline{A})=\frac{1}{V_N}\int_{CP^N}dv\overline{A}\ .\label{av}
\end{equation}

\bigskip
\bigskip
{\bf Acknowledgments}

I thank Ingemar Bengtsson for discussions.

\end{document}